\newcommand{\bscco}{BSCCO}
\newcommand{\opd}{OPD}
\newcommand{\sod}{SOD}
\newcommand{\ovd}{OVD}
\newcommand{\hod}{HOD}
\newcommand{\brg}{BrG}
\newcommand{\rsb}{RSB}
\newcommand{\fo}{FO}
\newcommand{\so}{SO}
\newcommand{\llll}{LLL}
\newcommand{\gl}{GL}

\documentclass[prl,twocolumn,amsmath,amssymb,showpacs,noshowkeys,superscriptaddress]{revtex4}

\usepackage{graphicx}
\usepackage{dcolumn}
\usepackage{bm}
\usepackage{color}

\begin{document}

\title{Interplay of Anisotropy and Disorder in the Doping-Dependent\\Melting and Glass Transitions of Vortices in Bi$_2$Sr$_2$CaCu$_2$O$_{8+\delta}$}

\author{H.~Beidenkopf}
\email{haim.beidenkopf@weizmann.ac.il}
\author{T.~Verdene}
\author{Y.~Myasoedov}
\author{H.~Shtrikman}
\author{E.~Zeldov}
\affiliation{Department of Condensed Matter Physics, Weizmann
Institute of Science, Rehovot 76100, Israel}%
\author{B.~Rosenstein}
\affiliation{Electrophysics Department, National Chiao Tung University, Hsinchu 30050, Taiwan, Republic of China}%
\author{D.~Li}
\affiliation{Department of Physics, Peking University, Beijing 100871, China}%
\author{T.~Tamegai}
\affiliation{Department of Applied Physics, The University of Tokyo, Hongo, Bunkyo-ku, Tokyo 113-8656, Japan}%
\date{\today}

\begin{abstract}
We study the oxygen doping dependence of the equilibrium first-order
melting and second-order glass transitions of vortices in
Bi$_2$Sr$_2$CaCu$_2$O$_{8+\delta}$. Doping affects both anisotropy
and disorder. Anisotropy scaling is shown to collapse the melting
lines only where thermal fluctuations are dominant. Yet, in the
region where disorder breaks that scaling, the glass lines are still
collapsed. A quantitative fit to melting and replica symmetry
breaking lines of a 2D Ginzburg-Landau model further reveals that
disorder amplitude weakens with doping, but to a lesser degree than
thermal fluctuations, enhancing the relative role of disorder.

\end{abstract}



\maketitle

Elasticity, thermal energy, disorder, and inter-layer coupling are
some of the closely competing energy scales in the intricate $H-T$
phase diagram of the vortex matter in the layered high temperature
superconductor Bi$_2$Sr$_2$CaCu$_2$O$_{8+\delta}$ (\bscco{})
\cite{ertas.d.pc272,mikitik.gp.prb68,kierfeld.j.prb69,giamarchi.t.prb55,vinokur.v.pc295,glazman.li.prb43}.
The low-temperature part of the equilibrium phase diagram of
\bscco{} was recently made accessible to experiment by vortex
shaking \cite{willemin.m.prl81,avraham.n.nat411,beidenkopf.h.prl95}.
It unveiled a first-order (\fo{}) inverse melting line, which
continues the thermal melting line from high temperatures
\cite{avraham.n.nat411} separating low-field ordered phases from
high-field amorphous ones. A second-order (\so{}) transition line,
at which low-temperature glassy phases get thermally depinned, was
subsequently reported \cite{beidenkopf.h.prl95}. In this letter we
study the oxygen doping dependence of these transition lines. We
show that the \so{} line scales with material anisotropy even where
the \fo{} line does not, and that effective disorder weakens with
doping, but gains relative dominance over thermal fluctuations.

We present measurements of optimally doped (\opd{}), slightly over
doped (\sod), over doped (\ovd{}) and highly over doped (\hod{})
\bscco{} crystals \cite{ooi.s.pc302, motohira.n.jcsp97} with
critical temperatures $T_c=$92, 90, 88.5 and 86 K, respectively,
corresponding to hole concentrations of 0.171, 0.180, 0.184 and
0.190 \cite{kim.gc.prb72}. Various crystal geometries were studied
with typical sizes of $\sim$300$\times$300$\times$40 $\mu$m$^3$.
They were mounted on $10\times10$ $\mu$m$^{2}$ Hall sensor arrays,
fabricated in a GaAs/AlGaAs heterostructure. At low temperatures we
utilized a 350 Oe in-plane \textsl{ac} shaking field of 10 Hz to
relax the pancake vortices towards their equilibrium configuration
\cite{willemin.m.prl81,avraham.n.nat411,beidenkopf.h.prl95}.
Conjugating local probes with shaking yielded the equilibrium
reversible magnetization of the samples.

\begin{figure} [!t] \centering
\mbox{\includegraphics[width=0.49\textwidth]{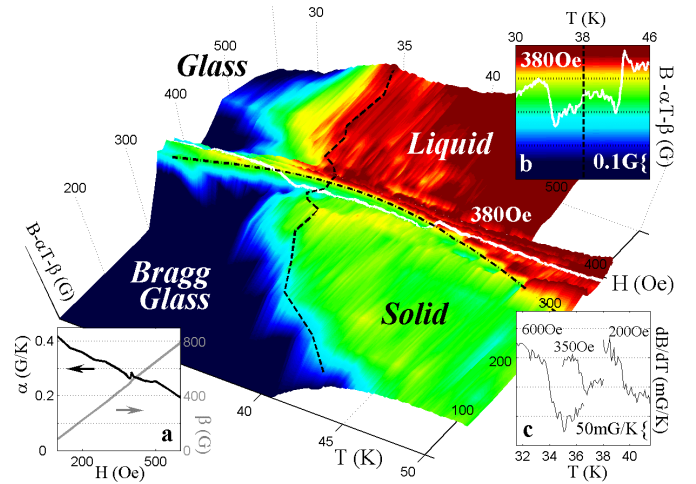}}
\caption{(color online) The local induction shows \fo{}
(dashed-dotted) and \so{} (dashed) transition lines in the $H-T$
phase diagram. The induction in the high-temperature depinned liquid
and solid phases was artificially flattened by a linear subtraction
$\alpha_{_H} T + \beta_{_H}$ (hence the constant color). The
low-temperature \brg{} and glassy phases have thus a finite positive
slope. (\textbf{a}) Values of $\alpha_{_H}$ and $\beta_{_H}$ used.
(\textbf{b}) FO-SO-FO transition sequence at a 380 Oe temperature
sweep on the background of the colorbar. (\textbf{c}) Discontinuous
steps in $dB/dT$, manifesting the \so{} nature of the transition.}
\label{fig1}
\end{figure}
Figure \ref{fig1} shows the local induction, $B(H,T)$, measured in
the \sod{} sample by sweeping the temperature, $T$, at a constant
out-of-plane field, $H$, in presence of an in-plane shaking field. A
linear term $\alpha_{_H} T+ \beta_{_H}$ was subtracted from each
temperature sweep (Fig.~\ref{fig1}a) to flatten the originally
increasing ($\alpha_{_H}>0$) local induction. As a result, the \fo{}
and \so{} singular behavior can be readily traced throughout the
$H-T$ phase diagram.

The \fo{} melting transition is manifested by a discontinuous step
in the magnetization along the dashed-dotted line in
Fig.~\ref{fig1}. It separates the high-field amorphous glass and
liquid phases from the low-field quasi-long-range-ordered Bragg
glass (\brg{}) \cite{zeldov.e.nat375,giamarchi.t.prb55}. The melting
line becomes nonmonotonic at a certain temperature
\cite{avraham.n.nat411, beidenkopf.h.prl95}. Its inverse-melting
part is believed to be an order-disorder transition, induced mainly
by quenched disorder and not by thermal fluctuations
\cite{ertas.d.pc272, khaykovich.b.prb56,
kierfeld.j.prb69,giamarchi.t.prb55,vinokur.v.pc295,
mikitik.gp.prb68, radzyner.y.prb65, olsson.p.prl87}.

The \so{} phase transition is manifested by a break in the slope of
the magnetization at $T_\mathrm{g}$, marked in Fig.~\ref{fig1} by
the dashed line. It separates the fully flattened high-temperature
magnetization ($dB/dT|_{T>T_\mathrm{g}} - \alpha \approx 0$) of
constant color from a low-temperature region of nonzero slope
($dB/dT|_{T<T_\mathrm{g}} - \alpha > 0$), whose color varies with
temperature. The nonanaliticity at $T_\mathrm{g}$ is demonstrated in
Fig.~\ref{fig1}c, which shows $\sim$40 mG/K steps in $dB/dT$ for
temperature sweeps lying below as well as above the \fo{} line. The
location of the \so{} line above melting resembles earlier dynamic
irreversibility measurements \cite{cubitt.r.nat365, zeldov.e.epl30,
konczykowski.m.pc332, gaifullin.mb.prl84, portier.f.prb66} and
theoretical models of a glass transition \cite{ertas.d.pc272,
vinokur.v.pc295,giamarchi.t.prb55, kierfeld.j.prb69,
mikitik.gp.prb68, nonomura.y.prl86, ryu.s.prl77,rodriguez.jp.prb73}.
Few dynamic measurements found a similar line of thermal depinning
below melting \cite{fuchs.dt.prl80, beidenkopf.h.prl95,
dewhurst.cd.prb56, matsuda.y.prl78, ooi.s.pc378, sugano.r.pc388,
yamaguchi.y.prb63}, which separates the \brg{} phase below it from a
thermally depinned solid, whose detailed characteristics are not yet
certain.

\begin{figure} [!t] \centering
\mbox{\includegraphics[width=0.49\textwidth]{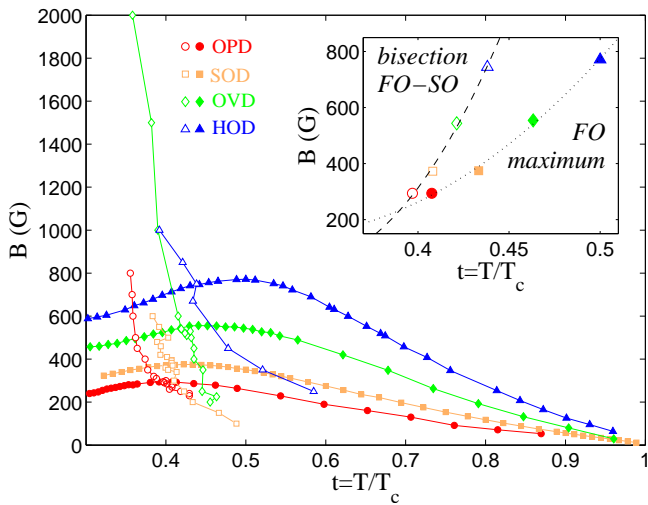}}
\caption{(color online) \fo{} and \so{} transition lines (solid and
open symbols, respectively) measured with \opd{}
({\Large{$\circ$}}), \sod ({$\square$}), \ovd{} ({$\lozenge$}{}) and
\hod{} ({$\vartriangle$}) samples. The inset shows the doping
dependence of the maximum of the \fo{} line and its bisection with
the \so{} line. Dotted and dashed lines are second-degree polynomial
fits.} \label{fig2}
\end{figure}

We mapped three more samples of various doping levels that yielded
the $B-T$ phase diagram plotted in Fig.~\ref{fig2}. The melting line
shifts as a whole to higher fields with doping due to the decrease
in anisotropy, which implies stronger inter-layer coupling and
higher stiffness of the pancake vortex stacks
\cite{khaykovich.b.prl76, hanaguri.t.pc256, ooi.s.pc302,
yamaguchi.y.prb63}. As the \fo{} line shifts to higher fields, its
maximum shifts towards increasingly higher temperatures
(Fig.~\ref{fig2} inset). Identifying these maxima with the crossover
from disorder to thermally dominated behavior \cite{ertas.d.pc272,
kierfeld.j.prb69, mikitik.gp.prb68,giamarchi.t.prb55,
vinokur.v.pc295} may suggest that doping enhances disorder. The
\so{} transition line also shifts upwards, further signifying that
disorder is enhanced with doping. However, we will show below that
this conclusion is inaccurate. Moreover, arguing naively that the
balance between characteristic energies constrains the \so{} line to
bisect the \fo{} one at its maximum \cite{ertas.d.pc272} is also too
simplistic. The inset of Fig.~\ref{fig2} clearly shows that in
over-doped samples the bisection point resides below the \fo{}
maximum temperature (the overlap of the two in the \opd{} sample is
apparently accidental). Therefore, a naive description of thermal
depinning cannot account for both the \so{} line and the \fo{}
maximum behavior (for instance, the latter is directly affected also
by elasticity) \cite{vinokur.v.pc295, mikitik.gp.prb68,
kierfeld.j.prb69}.

We now present a quantitative scaling analysis of the doping
dependence, parameterized by the sample anisotropy ratio
$\varepsilon^2 = m_\mathrm{ab}/m_\mathrm{c}$ ($m_i$ is the
electronic effective mass in the $i$th direction). For high-$\kappa$
superconductors, such as \bscco{}, the Ginzburg-Lanadau (\gl{}) free
energy functional can be recast into an isotropic ($\varepsilon=1$)
form by rescaling its parameters \cite{blatter.g.prl68}. We focus on
one such transformation \cite{klemm.ra.prb21} that rescales space,
magnetic induction, the penetration depth $\lambda_0$ and the
coherence length $\xi_0$. Models for high-temperature melting are
usually independent of $\xi_0$, and find $\lambda_0$ to enter with
some model-dependent power as a proportionality factor. For
definiteness we resort to a specific model \cite{daemen.ll.prl70,
glazman.li.prb43} for a \fo{} evaporation line with no disorder from
vortex solid to pancake gas $B_\mathrm{E}(t)\propto (\varepsilon^2
/\lambda_0^2 d)(1-t^2)/t$, where $t=T/T_c$ and $d$ is the
inter-plane separation. The scaling transformation
\cite{klemm.ra.prb21} indeed renders $B_\mathrm{E}(t)$ isotropic.

\begin{figure} [!t]
\centering \mbox{\includegraphics[width=0.49\textwidth]{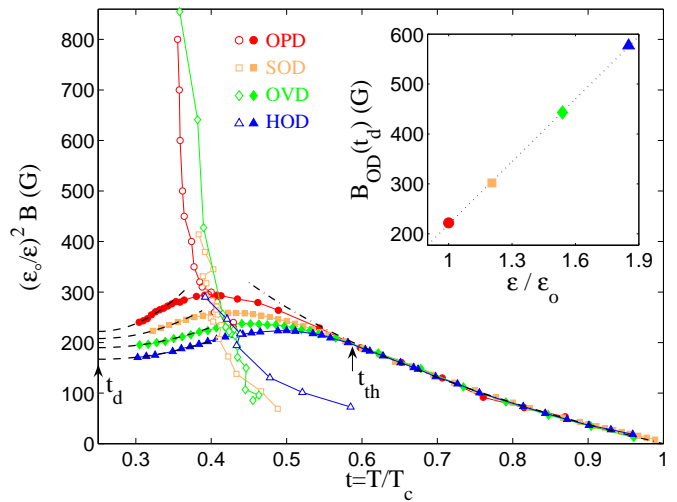}}
\caption{(color online) Data collapse of the over-doped
high-temperature ($t>t_\mathrm{th}$) \fo{} lines (solid symbols)
onto the \opd{} line, which also collapses the SO lines (open
symbols). The fit to $B_\mathrm{E}(t)$ (dashed-dotted) sets the
anisotropy ratios $\varepsilon/\varepsilon_\circ$ to 1, 1.20, 1.54
and 1.85 for the \opd{} ({\Large{$\circ$}}), \sod{} ({$\square$}),
\ovd{} ({$\lozenge$}) and \hod{} ({$\vartriangle$}) samples,
respectively. The inset shows the linear dependence of the
characteristic $B_\mathrm{OD}(t_\mathrm{d})$ on
$\varepsilon/\varepsilon_\circ$ (dotted line).} \label{fig3}
\end{figure}

Figure \ref{fig3} shows the rescaled \fo{} and \so{} lines. The
high-temperature parts of the \fo{} lines are perfectly collapsed by
dividing their induction axes by a constant
\cite{yamaguchi.y.prb63}. $B_\mathrm{E}(t)$ (dashed-dotted line)
 fits the collapsed melting lines precisely, asserting that the
multiplicands in this procedure are
$(\varepsilon_\circ/\varepsilon)^2$, normalized by the \opd{}
$\varepsilon^{-1}_\circ \approx 500$ \cite{tokunaga.m.prb67}.
Anisotropy scaling collapses the data down to $t_\mathrm{th} \approx
0.58$. Below $t_\mathrm{th}$, which appears to be independent of
anisotropy, the rescaled \fo{} lines disperse again, and the fit to
$B_\mathrm{E}(t)$ breaks. The flattening of the \fo{} line towards
an inverse-melting behavior results from quenched disorder, which
gains dominance with decreasing temperatures \cite{ertas.d.pc272}.
Accordingly, above $t_\mathrm{th}$ the \fo{} transition is purely
thermally-induced, and completely unaffected by disorder
\cite{khaykovich.b.prb56}. Just below $t_\mathrm{th}$ disorder
becomes a relevant, though not yet a dominant, energy scale.

This counterparts the extremely low-temperature behavior, where
thermal energy becomes negligible relative to pinning, resulting in
a flat temperature-independent behavior of the \fo{} lines
\cite{ertas.d.pc272}. Indeed the \fo{} lines in Fig.~\ref{fig3} tend
to flatten towards their ends, below which 350 Oe - 10 Hz shaking is
insufficient for detecting a reversible melting step. We thus
conjectured a similar doping-independent threshold temperature
$t_\mathrm{d} \approx 0.25$, below which thermal energy becomes
irrelevant. We fitted the low-temperature order-disorder lines by a
leading order expansion $B_\mathrm{OD}(t \gtrsim t_\mathrm{d}) \sim
B_\mathrm{OD}(t_\mathrm{d}) + \Lambda (t-t_\mathrm{d})^2$, which
agrees very well with measurement. We can, therefore, estimate
$B_\mathrm{OD}(t_\mathrm{d})$, at which elasticity is balanced
solely by the disordering potential. It increases monotonically with
doping (inset of Fig.~\ref{fig3}), stating that with reducing
anisotropy elasticity gains dominance also over disorder.

Yet, the most remarkable outcome of the anisotropy scaling shown in
Fig.~\ref{fig3} is the simultaneous collapse of the \so{} transition
lines (with zero freedom). The \so{} lines reside in a temperature
region where anisotropy scaling of the melting lines fails due to
effects of disorder. Still, the same scaling transformation somehow
succeeds in rendering the glass-transition isotropic, even though
this transition is believed to be intimately related with the
competition between disorder and thermal fluctuations.

To gain deeper understanding of the low-temperature behavior we fit
the measured transition lines to those predicted by a recent
calculation \cite{digping.l.abr80}, which gives access to the doping
dependence of the model's free parameters. It incorporates thermal,
disordering and elastic energies to yield bisecting \fo{} and \so{}
lines. The pancake vortex system is modeled by the 2D \gl{} theory.
We interpret this single layer model as the outcome of an
integration of all other layers out of a complete 3D theory,
resulting in an effective 2D model whose renormalized coefficients
may still depend on anisotropy $\varepsilon$ of a 3D mass tensor.

The free energy functional is averaged over gaussian disorder in the
coefficients of the quadratic and quartic terms using the replica
method. Taking the lowest-Landau-level (\llll{}) approximation
yields for the replicated partition function
\begin{eqnarray*}
\overline{\mathcal{Z}^n} = \int_{\Psi_1 \ldots \Psi_n} \exp \bigg(
-\sum_a G_0(\Psi_a)+ \sum_{a,b} \tilde{R} |\Psi_a|^2 |\Psi_b|^2
\bigg),\\
G_0(\Psi) = \int \frac{d^2x}{4\pi} \left( a_T |\Psi |^2 +
\frac{|\Psi |^4}{2} + \frac{\kappa^2 (b-h)^2}{\pi \sqrt{2 Gi} b t}
\right) ,\quad
\end{eqnarray*}
where $\tilde{R} = a_T^2 bR / 32\pi^2$, $R$ is the disorder
amplitude in the coefficient of the quadratic \llll{} term,
$b=B/\tilde{H}_{c2}$, $a_T = (b+t-1)/(2\pi^2 b^2t^2Gi)^{1/4}$ is the
\llll{} parameter, $Gi$ is the effective Ginzburg number that
generally scales thermal fluctuations, and $\kappa$ is the \gl{}
parameter. Randomness in the coefficient of the quartic \llll{}
term, although crucial for obtaining the non-analyticity of \so{},
has negligible effect on the locus of the transition lines, and is
therefore neglected in the present context.

The \fo{} line is calculated by equating the energies of the
homogeneous and crystalline states under the influence of disorder
in gaussian approximation. This extends beyond the earlier
calculation, which treated disorder perturbatively
\cite{li.d.prl90}. The glass line is found from the stability
analysis of the replica symmetric solution. The replica symmetry
breaking (\rsb{}) is continuous. The corresponding Parisi function
describing the hierarchial structure of the glassy state and its
detailed derivation can be found in Ref.~\cite{digping.l.abr80}.

\begin{figure} [!t]
\centering \mbox{\includegraphics[width=0.49\textwidth]{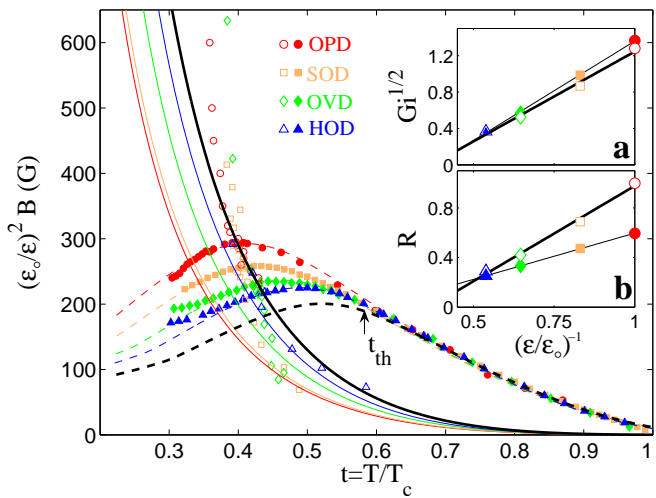}}
\caption{(color online) Fits of \fo{} (dashed) and \so{} (solid)
lines of the \llll{} 2D \gl{} model with $\tilde{H}_{c2}=$100 T,
$\tilde{T}_c=$103, 101, 98 and 95 K for the \opd{}
({\Large{$\circ$}}), \sod{} ({$\square$}), \ovd{} ({$\lozenge$}{})
and \hod{} ({$\vartriangle$}) samples, respectively. The values of
$Gi$ (\textbf{a}) and $R$ (\textbf{b}) were fixed either by fitting
the \so{} lines and \fo{} ones above $t_{\rm{th}}$ (thick lines) or
by fitting the full \fo{} lines (thin lines).}\label{fig4}
\end{figure}

The model's free parameters are $R$, $Gi$, $\tilde{T}_c$, and
$\tilde{H}_{c2}=T_c \frac{dH_{c2}}{dT}|_{_{T_c}}$. However, due to a
hidden symmetry, given by $Gi/\tilde{H}_{c2}^2$ and $Gi/R^2$, the
theory effectively has only three independent fitting parameters. We
therefore fixed $\tilde{H}_{c2}=$100T, for simplicity, consistent
with our limited range of hole concentrations \cite{kim.gc.prb72}.
Figure \ref{fig4} presents two fitting strategies motivated by
different physical behaviors. The first (thick lines) is optimized
to fit the collapsed parts of the phase diagram - \so{} lines and
\fo{} ones for $t>t_{\rm{th}}$. Interestingly, this set
simultaneously provides a collapsed \fo{} line also for
$t<t_{\rm{th}}$. This results from the above symmetry of the model
in which $Gi$ and $\varepsilon$ take the same role, defining a
relation $R\propto \sqrt{Gi}$ along which the calculated phase
diagram remains unchanged under anisotropy rescaling. The thick
lines in Fig.\ref{fig4} insets show the resulting parameters
$Gi=3.91(\varepsilon_0/\varepsilon-0.37)^2$ and
$R=1.54(\varepsilon_0/\varepsilon-0.37)$. The departure of the
calculated \fo{} line (thick dashed) from the measured ones below
$t_{\rm{th}}$ may indicate that theory lacks a symmetry-breaking
mechanism once anisotropy scaling breaks in experiment.

An alternative fit (thin lines), faithful to the complete \fo{}
lines, dictates the slightly different set of values
$Gi=4.77(\varepsilon_\circ/\varepsilon-0.38)^2$, $R =
0.75(\varepsilon_\circ/\varepsilon-0.2)$ (insets, thin lines). The
excellent fit of the \fo{} lines (dashed) suggests that the
effective 2D model still captures the essential physics involved.
Its validity should break down at very low temperatures, where
indeed it misses the flattening of the \fo{} lines as $t_\mathrm{d}$
is approached, and close to $T_c$ due to critical fluctuations,
which may explain the $10 \%$ overestimate in the fitted bare
$\tilde{T_c}$ values \cite{larkin.a.fluctsinscs}. Substituting this
set of values in the calculation of the \rsb{} lines (solid) with
zero freedom also produces a good fit that improves with doping, but
misses the anisotropy rescaling of the measured \so{} lines.
Nevertheless, the discrepancy between the two sets of values of the
different fitting strategies is small, and vanishes in the high
doping limit.

In Fig.~\ref{fig4}a the reduced dominance of thermal fluctuations
with doping is clearly captured by the decrease in $Gi$. Its
dependence on anisotropy is closer to the 3D $1/\varepsilon^2$ than
to the 2D $\varepsilon$-independent behavior. Remarkably, the
disorder amplitude in Fig.~\ref{fig4}b also decreases with doping.
This contrasts the result of Ref.~\cite{li.tw.pc257}, which found
that oxygen addition increases the defects concentration. The
apparent contradiction is resolved by noting that the disorder
amplitude is affected also by the anisotropy dependence of the
effective pinning potential, which decreases with doping as the
vortex stacks get stiffer. The overall decrease of $R$ suggests that
the latter mechanism is dominant, not the rising defect
concentration.

Last, while both $Gi$ and $R$ decrease with doping, $Gi$ has a
stronger dependence on anisotropy. This clarifies the true mechanism
that shifts the maximum of the \fo{} line and its bisection with the
\so{} line towards higher temperatures with doping
(Fig.~\ref{fig2}). Although the magnitude of the disorder amplitude
decreases with increased doping, disorder still gains
\emph{relative} dominance over thermal fluctuations, which decrease
faster with $\varepsilon$.

In summary,  $\varepsilon^2$ scaling accounts for the reduced
anisotropy of \bscco{} samples with doping down to $t_\mathrm{th}$.
Below $t_\mathrm{th}$, where disorder becomes relevant,
$\varepsilon^2$ scaling still collapses the \so{} lines. From the
quantitative agreement with the \llll{} 2D \gl{} model we deduce
that disorder and thermal-fluctuations weaken relative to elasticity
with increased doping, which shifts the \fo{} line towards higher
fields. Disorder still gains relative dominance over thermal
fluctuations, which concurrently shifts the maximum of the \fo{}
line and its bisection with the \so{} line towards higher
temperatures.

We thank V.M.~Vinokur, and G.P.~Mikitik for stimulating discussions.
This work was supported by the Israel Science Foundation Center of
Excellence, by the German-Israeli Foundation G.I.F., by Grant-in-aid
for Scientific Research from the Ministry of Education, Culture,
Sports, Science, and Technology, Japan, by the National science
foundation of China, and by the MOE ATU Program of R.O.C.
NSC952112M009048. BR acknowledges the support of the Albert Einstein
Minerva Center for Theoretical Physics and EZ the US-Israel
Binational Science Foundation (BSF).


\end{document}